\documentclass
[floats,superscriptaddress,balancelastpage,prl,a4paper,twocolumn,tightenlines,showpacs,10pt]{revtex4}%
\usepackage{graphicx}
\usepackage{amsmath}
\usepackage{amsfonts}
\usepackage{amssymb}%
\begin{document}
\title{Experimental realization of optimal asymmetric cloning and telecloning via
partial teleportation}
\author{Zhi Zhao}
\affiliation{Hefei National Laboratory for Physical Sciences at Microscale and Department
of Modern Physics, University of Science and Technology of China, Hefei, Anhui
230027, People's Republic of China}
\affiliation{Department of Microelectronics and Information Technology, Royal Institute of
Technology (KTH), S-16440 Kista, Sweden}
\author{An-Ning Zhang}
\affiliation{Hefei National Laboratory for Physical Sciences at Microscale and Department
of Modern Physics, University of Science and Technology of China, Hefei, Anhui
230027, People's Republic of China}
\author{Xiao-Qi Zhou}
\affiliation{Hefei National Laboratory for Physical Sciences at Microscale and Department
of Modern Physics, University of Science and Technology of China, Hefei, Anhui
230027, People's Republic of China}
\author{Yu-Ao Chen}
\affiliation{Hefei National Laboratory for Physical Sciences at Microscale and Department
of Modern Physics, University of Science and Technology of China, Hefei, Anhui
230027, People's Republic of China}
\author{Chao-Yang Lu}
\affiliation{Hefei National Laboratory for Physical Sciences at Microscale and Department
of Modern Physics, University of Science and Technology of China, Hefei, Anhui
230027, People's Republic of China}
\author{Anders Karlsson}
\affiliation{Department of Microelectronics and Information Technology, Royal Institute of
Technology (KTH), S-16440 Kista, Sweden}
\author{Jian-Wei Pan}
\affiliation{Hefei National Laboratory for Physical Sciences at Microscale and Department
of Modern Physics, University of Science and Technology of China, Hefei, Anhui
230027, People's Republic of China}
\affiliation{Physikalisches Institut, Universit\"{a}t Heidelberg, Philosophenweg 12, 69120
Heidelberg, Germany}
\date{\today }

\begin{abstract}
We report an experimental realization of both optimal asymmetric cloning and
telecloning of single photons by making use of partial teleportation of an
unknown state. In the experiment, we demonstrate that, conditioned on the
success of partial teleportation of single photons, not only the optimal
asymmetric cloning can be accomplished, but also one of two outputs can be
transfered to a distant location, realizing the telecloning. The experimental
results represent a novel way to achieve the quantum cloning and may have
potential applications in the context of quantum communication.

\end{abstract}

\maketitle

Quantum cloning is a process to distribute the quantum information in a state
onto multiple output states. However, perfect quantum cloning of an unknown
state is forbidden due to the restriction of quantum no-cloning theorem
\cite{nocloning}. To understand the underlying limits, Bu\v{z}ek and Hillery
first proposed a symmetric \emph{universal quantum-copying machine
}(UQCM)\emph{\ }\cite{buzek}\emph{\ }that produces two identical outputs whose
quality is independent of the input states. The UQCM was later proved to be
optimal \cite{gisin1,bruss1}, and constitutes the optimal attack in the
six-state protocol of quantum cryptography \cite{bruss2, gisin2}.

Because of its fundamental importance in quantum mechanics and quantum
cryptography, various UQCMs have been demonstrated either by constructing the
complex quantum network \cite{huang,cum}, or by exploiting the process of
stimulation emission \cite{simon,dik01,gisin3,marti04} or by photon bunching
\cite{marti05,dik00}. However, in all these experiment, only identical outputs
have been realized.

To investigate the asymmetric distribution of an unknown quantum state, Cerf
\cite{cerf} first proposed a family of Pauli cloning machines that produced
two unnecessarily identical output qubits. The state-independent fidelities
$F_{e}$ and $F_{d}$ of the two copies was bound to a no-cloning inequality
\cite{cerf}%

\begin{equation}
\left(  1-F_{d}\right)  \left(  1-F_{e}\right)  \geq\left[  1/2-\left(
1-F_{d}\right)  -\left(  1-F_{e}\right)  \right]  ^{2}.
\end{equation}
This inequality sets the optimal tradeoff between the quality of the two
copies in the sense that for given fidelity\textit{ }$F_{d}$ one cannot obtain
a better fidelity $F_{e}$ \cite{cerf,niu}.\textit{ }The optimal unbalanced
fidelities quantify the novel cloning limit imposed by quantum mechanics,
thereby extending the results on the symmetric UQCM.

Moreover, in the context of quantum cloning, it is highly desirable to
transfer one of the quantum cloning to a distant location, realizing the
so-called telecloning \cite{murao1,murao2}. Although it is, in principle,
straightforward to combine quantum teleportation with optimal quantum cloning,
it will involve the extra resources and also leads to the extremely low
efficiency with the existing technology. Alternatively, the procedure could be
reduced to exploit the particular multiparticle entangled states
\cite{murao1}, but such crucial resource has not yet been experimentally demonstrated.

In this letter, we report an experimental realization of both asymmetric
cloning and telecloning by making use of partial teleportation of an unknown
state \cite{filip}. In the experiment, we demonstrate that, conditioned on the
partial quantum teleportation of single photons, not only asymmetric cloning
can be realized, but also one of two outputs can be transfered to a distant
location, realizing the telecloning.%

\begin{figure}
[ptb]
\begin{center}
\includegraphics[
height=2.2182in,
width=2.9992in
]%
{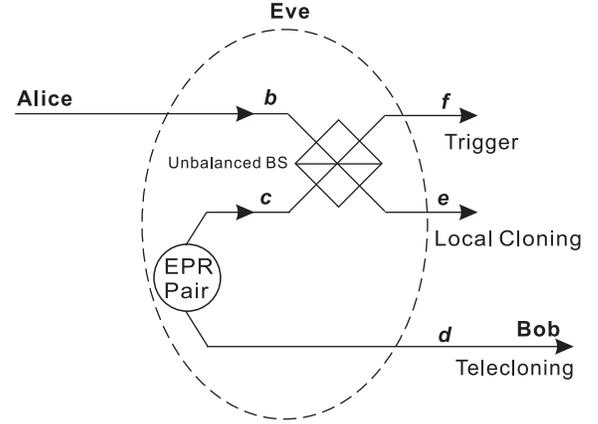}%
\caption{Scheme of asymmetric cloning and telecloning by making use of partial
teleportation.}%
\end{center}
\end{figure}

Let us first consider the scenario that Alice wants to send an unknown
polarization state of a single photon in a mode $b$ to Bob at a distant
location. Eve seeks to extract partial (or full) quantum information of the
state by using partial quantum teleportation \cite{bennett, dik02} having a
pair of entangled photons in modes $c$ and $d$, (Fig. 1). The procedure
\cite{filip} is that Eve performs a partial Bell-state measurement on the
photons in mode $b$ and $c$, then resend the teleported state to Bob in mode
$d$. The pair of entangled photon held by Eve is in the Bell state%

\begin{equation}
|\Psi^{-}\rangle_{cd}=\frac{1}{\sqrt{2}}(\left\vert H\right\rangle
_{c}\left\vert V\right\rangle _{d}-\left\vert V\right\rangle _{c}\left\vert
H\right\rangle _{d}),
\end{equation}
and the partial Bell state measurement is achieved through an unbalanced beam
splitter (BS) with a variable reflectivity, $0\leq R\leq0.5$.

Consider, for example, one vertically polarized photon is sent by Alice to
Bob. Then, in the above eavesdropping protocol, the evolution of the initial
state $\left\vert V\right\rangle _{b}\otimes|\Psi^{-}\rangle_{cd}$ is
determined by the evolution of photons in modes $b$ and $c$,%

\begin{equation}
b\rightarrow\left(  irf+te\right)  ,\text{ }c\rightarrow\left(  tf+ire\right)
,
\end{equation}
where $R=r^{2}$, $1-R=t^{2}$. If we restrict ourself to the cases where both
photons leave the beam splitter separately, then we obtain the following state (unnormalized)%

\begin{equation}
(t^{2}\left\vert V\right\rangle _{e}\left\vert H\right\rangle _{f}%
-r^{2}\left\vert H\right\rangle _{e}\left\vert V\right\rangle _{f})\left\vert
V\right\rangle _{d}-\left(  t^{2}-r^{2}\right)  \left\vert V\right\rangle
_{e}\left\vert V\right\rangle _{f}\left\vert H\right\rangle _{d}.
\end{equation}
Thus, tracing over photons in modes $d$ and $f$ and measuring the probability
of the output to be vertically polarized as the input photon, Eve could obtain
the local cloning in mode $e$ with a fidelity \cite{filip}%

\begin{equation}
F_{e}\left(  R\right)  =\frac{1}{2P\left(  R\right)  }\left[  \left(
1-2R\right)  ^{2}+\left(  1-R\right)  ^{2}\right]  ,
\end{equation}
and similarly Bob would obtain the telecloning state in mode $d$ with a
fidelity \cite{filip}:%

\begin{equation}
F_{d}\left(  R\right)  =\frac{1}{2P\left(  R\right)  }\left[  R^{2}+\left(
1-R\right)  ^{2}\right]  ,
\end{equation}
where $P\left(  R\right)  =1-3R+3R^{2}$,\textbf{ }corresponds to the
probability that the two photons exit Eve's beamsplitter seperately. Although
the above scheme only succeed probabilistically, it is suffcient to provide a
proof-in-principle demonstration of both the optimal asymmetric cloning and telecloning.

It follows \cite{cerf,niu,filip}, that the fidelities $F_{e}$ and $F_{d}$
saturate the no-cloning inequality (1) and represent the optimal asymmetric
distribution of the initial quantum information between the local clone and
the distant clone. For the reflectivity of R=1/3 the protocol \cite{filip} was
reduced to the symmetric distribution \cite{buzek} with fidelities
$F_{e}=F_{d}=5/6,$ but one of the two clones was distributed to a distant
location, realizing the telecloning. The foregoing analysis are justified for
any input polarization owing to the rotational invariance of the Bell state in
Eq. (1) \cite{dik01}.%

\begin{figure}
[ptb]
\begin{center}
\includegraphics[
height=2.4967in,
width=3.1946in
]%
{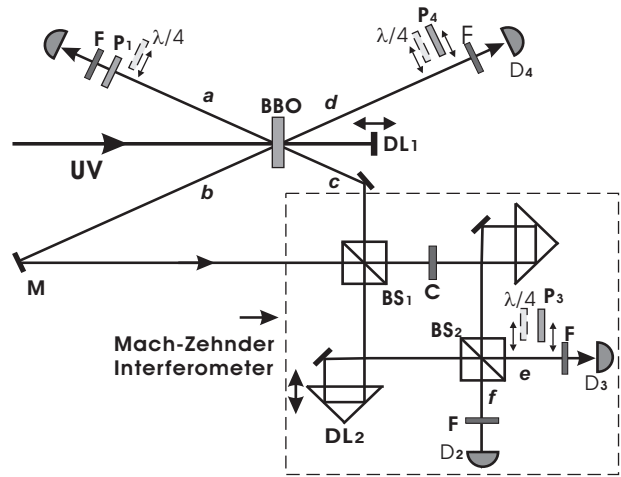}%
\caption{The schematic of the experimental apparatus used to demonstrate the
asymmetric cloning and telecloning. Two photons $b$ and $c$ out of entangled
pairs are first overlapped at the beam splitter (BS$_{1}$) and then recombined
at the BS$_{2}$ where the path lengths of the two photons have been adjusted
so that they arrive at two beam splitter simultaneously. The two BS together
with the compensator (C) constitute the two-photon Mach-Zehnder interferometer
\cite{rarity} that plays the role of the variable beam splitter. Polarizers
(P$_{1}$, P$_{3}$ and P$_{4}$) and $\lambda/4$ plates in front of the
detectors allow measurement of linear and circular R/L polarization. DL$_{1}$
is a delay mirror to change the delay between photons in modes $b$ and $c$,
and DL$_{2}$ is a prism to change the delay between the two arms of the
interferometer.}%
\end{center}
\end{figure}

A schematic of our experimental apparatus is shown in Fig. 2. We first
generate two pairs of entangled photons in the maximally entangled state
$|\Psi^{-}\rangle$ by type II down-conversion \cite{kwiat95}\ from an
ultraviolet (UV) pulsed laser in a BBO crystal. The UV pulse passing through
the crystal twice creates two pair of entangled photons in modes $a$-$b$ and
$c$-$d$. The UV pulsed laser with a central wavelength of 394nm has a pulse
duration of 200fs, a repetition rate of 76MHz, and an average power of 450mW.
Photons in modes $b$ and $c$ are first overlapped at the BS$_{1}$ and then
recombined at the BS$_{2}$ where the path lengths of the two photons have been
adjusted so that they arrive at two beam splitters simultaneously. Through
spectral filtering ($\Delta\lambda_{FWHM}=3$nm) \cite{marek95} and
fiber-coupled single-photon detectors, we can ensure that all the four photons
are in the perfect temporal and spatial mode overlap.

In the experiment, the crucial requirement is to overlap two photons in the
Mach-Zehnder (M-Z) interferometer \cite{rarity}, with which the controllable
phase difference will lead to the desired variable reflectivity. However, when
two photons with different polarizations pass through the same two arms of the
interferometer, they will usually experience unbalanced phase differences due
to the birefringent effect of the BS$_{1}$, the BS$_{2}$, and prisms (DL$_{2}$
and its counterpart in Fig. 2). To overcome this difficulty, we incorporate a
1.2mm type I LBO crystal as a compensator (C in Fig. 2) to vary the phase
shift in one arm of the interferometer so that the identical phase difference
could be reached by tilting the compensator.%

\begin{figure}
[ptb]
\begin{center}
\includegraphics[
height=4.9528in,
width=2.4042in
]%
{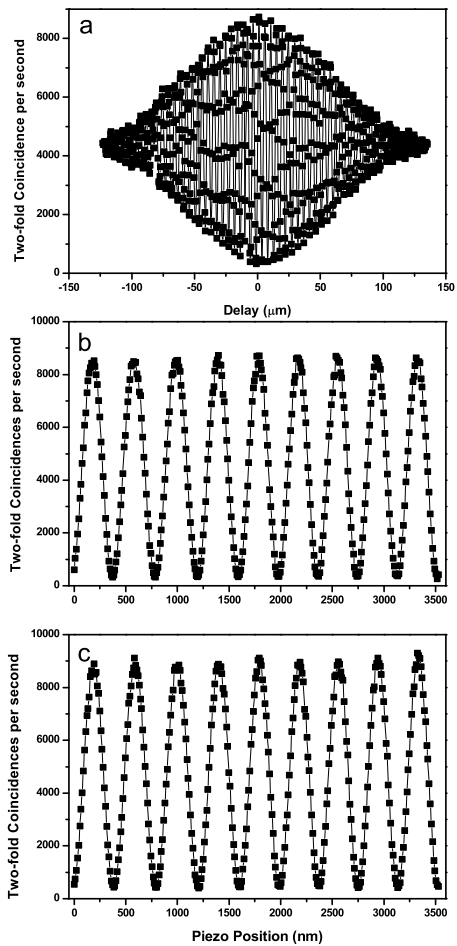}%
\caption{Experimental results showing that the two-photon Mach-Zehnder
interferometer works properly and exhibits the polarization-independent
reflectivity. In (a) and (b), we measure the twofold coincidence between the
modes $e$ and $a$ behind 0$^{0}$ polarizer. In (c), we measure the twofold
coincidence between the modes $f$ and $d$ behind 90$^{0}$ polarizer. (a), the
envelope of the observed twofold coincidences demonstrates that the single
photon passing through the interferometer interferes with itself. (b) and (c),
the two twofold coincidences exhibit a synchronized variation.This
demonstrates that the two photons with orthogonal polarizations have undergone
the identical phase difference, which consequently leads to the
polarization-independent reflectivity.}%
\end{center}
\end{figure}
%

\begin{figure}
[ptb]
\begin{center}
\includegraphics[
height=5.0851in,
width=2.3618in
]%
{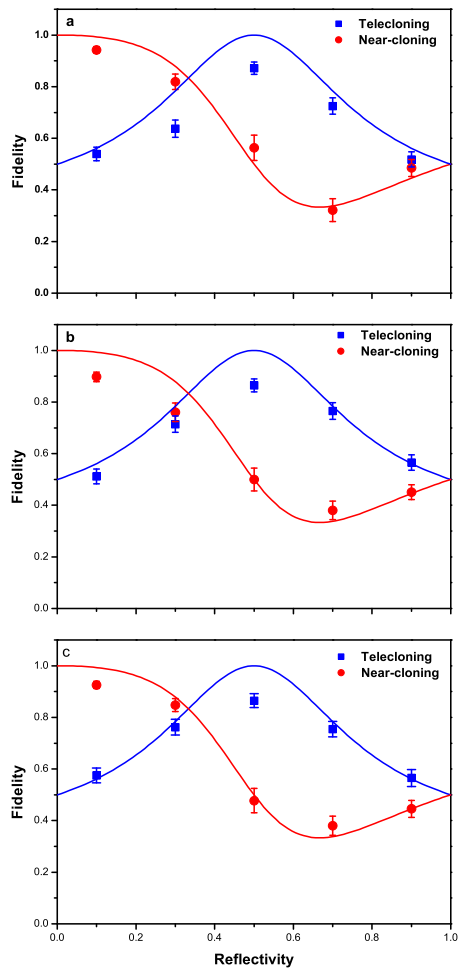}%
\caption{Experimental results demonstrating both the asymmetrical cloning and
telecloning of single photons in three complementary polarizations of 0$^{0}$
(a), +45$^{0}$ (b) and $L$ (c) with the various reflectivity beam splitter.
The experimental results are in well agreement with the theoretical prediction
(the solid plot) of Eq. (5) and (6).}%
\end{center}
\end{figure}

To show that the crucial M-Z interferometer works properly, two photons with
orthogonal polarizations in modes $b$ and $c$ are steered to the
interferometer. We predefine two photons in the vertical and horizontal
polarizations by performing the polarization measurements on the photons in
modes $a$ and $d$ behind 0$^{0}$ and 90$^{0}$ polarizers respectively. With
these settings, we first measure the twofold coincidence between the output
modes $a$-$e$, by scanning the position of the DL$_{2}$ in one arm of
interferometer with a step size of 0.36$\mu m$. The envelope of the observed
twofold coincidence demonstrates that the single photon interferes with itself
after passing through the interferometer (Fig. 3a). Then, we perform a fine
scan DL$_{2}$ around the centre of the envelope and simultaneously observe the
two independent twofold coincidences between the output modes $a$-$e$ and
modes $d$-$f$ in order to verify the polarization-independent reflectivity. We
slightly tilt the compensator inside the M-Z interferometer until the two
twofold coincidences exhibit a synchronized variation (Fig. 3b and 3c). This
demonstrates that the two photons of the orthogonal polarizations passing
through the interferometer have undergone the identical phase difference,
which consequently leads to the polarization-independent reflectivity. Further
results show that our compensation method also works for two photons in the
general polarizations.

Specifically, we obtained the reflectivity of R=1/2 when the phase difference
was set to be $\pi/2$. By scanning the DL$_{1}$, the perfect temporal overlap
was verified through a successful teleportation of photons polarized at
+45$^{0}$ with a visibility of $0.75\pm0.05$ at zero delay \cite{dik02}.

To further demonstrate our scheme, we varied the phase difference to achieve
various reflectivities of 0.1, 0.3, 0.5, 0.7, 0.9. The input photon (to be
cloned) was prepared in three complementary polarizations of 0$^{0}$,
+45$^{0}$ and circular left-handed, $L$ by triggering the polarization
measurement on the photon in mode $a$. Then the clones were further verified
by performing the conditional projection measurement on photons in modes $e$
and $d$. For example, conditioned on the detection of photon in mode $a$ as
well as the trigger measurements in modes $d$ and $f$, the local clones were
left in mode $e$, which were confirmed by performing the polarization
projection measurements. The fidelites were accordingly obtained by measuring
the probability of the output states to be in the input states. Similarly, we
could obtain the fidelities of the telecloning in mode $d$.

In our experiment, the integration time for each cloning measurement is 5
minutes while each reflectivity can vary less than 0.025. All these results
are shown in Fig. 4. From the figures, it is evident that the experimental
results are in well agreement with the theoretical prediction of Eq. (5) and
(6), while only $0\leq R\leq0.5$ represents the optimal asymmetrical cloning
and telecloning. The imperfect fidelities are mainly due to the instability of
the interferometer as well as the imperfections of the down-conversion source,
and the mode overlap of the photons inside the interferometer.

The experimental realization of both the optimal asymmetrical cloning and
telecloning\ deserves some further comments. First, in the experiment, an
unknown state was encoded into two qubit with ancilla pair of entangled photon
via partial quantum teleportation, which is significantly different from the
previous implementations
\cite{huang,cum,simon,dik01,gisin3,marti04,marti05,dik00}, and thus represents
a novel way to realize the quantum cloning. Second, both cloning and
teleportation are intrinsically integrated together so that it no longer
requires the extra resource to achieve the telecloning and will not lead to
any depressed efficiency, comparing with the Innsbruck experiment
\cite{dik02}. Third, although our present experimental demonstration required
the coincidence detection of all four photons, the telecloning could be freely
transferred to a distant location by using the nonpostselection teleportation
technique \cite{pan1}. For example, we could attenuate the photon intensity in
mode $b$ to suppress those spurious $e$-$f$ coincidence events, (i.e. those
events contributed by double pair emission either in modes $a$ and $b$ or $c$
and $d$), the threefold coincidence among the modes $a$, $e$ and $f$ would
then be sufficient to guarantee the success of partial teleportation. Thus,
the telecloning can be successfully achieved without the need to destructively
detect it.

In summary, we have for the first time presented an experimental realization
of both optimal asymmetrical cloning and telecloning via conditional partial
teleportation of an unknown state \cite{cerf,filip}. The experimental results
represent a novel way to achieve the quantum cloning and may have potential
applications in the context of future quantum communication.

\begin{acknowledgments}
This work was supported by the NSF of China, the CAS, the National Fundamental
Research Program, and the Swedish Foundation for Strategic Research-SSF INGVAR grant.
\end{acknowledgments}

\end{document}